\documentclass[aps,prl,floats,twocolumn,showpacs,superscriptaddress]{revtex4-1}   
\usepackage[dvips]{graphicx}
\usepackage{epsfig}
\usepackage{subfigure}
\usepackage{amsmath}
\usepackage{amssymb}
\usepackage{color}

\begin{document}

\title{Route towards classical frustration and band flattening via optical lattice distortion}
\author{Pil Saugmann}
\affiliation{Department of Physics, Stockholm University, Se-106 91
  Stockholm, Sweden} 
  \author{Jos\'e Vargas}
\affiliation{Institut f\"ur Laserphysik, Universit\"at Hamburg, 22761 Hamburg, Germany}
\affiliation{Zentrum f\"ur Optische Quantentechnologien and Institut f\"ur Laser-Physik, Universit\"at Hamburg, 22761 Hamburg, Germany}
  \author{Yann Kiefer}
\affiliation{Institut f\"ur Laserphysik, Universit\"at Hamburg, 22761 Hamburg, Germany}
\affiliation{Zentrum f\"ur Optische Quantentechnologien and Institut f\"ur Laser-Physik, Universit\"at Hamburg, 22761 Hamburg, Germany} 
  \author{Max Hachman}
\affiliation{Institut f\"ur Laserphysik, Universit\"at Hamburg, 22761 Hamburg, Germany}
\affiliation{Zentrum f\"ur Optische Quantentechnologien and Institut f\"ur Laser-Physik, Universit\"at Hamburg, 22761 Hamburg, Germany}
\author{Raphael Eichberger}
\affiliation{Institut f\"ur Laserphysik, Universit\"at Hamburg, 22761 Hamburg, Germany}
\affiliation{Zentrum f\"ur Optische Quantentechnologien and Institut f\"ur Laser-Physik, Universit\"at Hamburg, 22761 Hamburg, Germany} 
\author{Andreas Hemmerich}
\affiliation{Institut f\"ur Laserphysik, Universit\"at Hamburg, 22761 Hamburg, Germany}
\affiliation{Zentrum f\"ur Optische Quantentechnologien and Institut f\"ur Laser-Physik, Universit\"at Hamburg, 22761 Hamburg, Germany}
\affiliation{The Hamburg Center for Ultrafast Imaging, Luruper Chaussee 149, 22761 Hamburg, Germany}
\author{Jonas Larson}
\affiliation{Department of Physics, Stockholm University, Se-106 91
  Stockholm, Sweden} 
  \date{\today}

\begin{abstract}
We propose and experimentally explore a method for realizing frustrated lattice models using a Bose-Einstein condensate held in an optical square lattice. A small lattice distortion opens up an energy gap such the lowest band splits into two. Along the edge of the first Brillouin zone for both bands a nearly flat energy-momentum dispersion is realized. For the excited band a highly degenerate energy minimum arises. By loading ultracold atoms into the excited band, a classically frustrated $XY$ model is formed, describing rotors on a square lattice with competing nearest and next nearest tunnelling couplings. Our experimental optical lattice provides a regime, where a fully coherent Bose-Einstein condensate is observed, and a regime where frustration is expected. If we adiabatically tune from the condensate regime to the regime of frustration, the momentum spectra shows a complete loss of coherence. Upon slowly tuning back to the condensate regime, coherence is largely restored. Good agreement with model calculations is obtained. 
\end{abstract}
\maketitle

{\it Introduction.} -- Classically frustrated systems are characterized by highly degenerate ground states which may result in glassy or liquid like phases with exotic properties~\cite{frust1,spinglass,spinliq}. While extensively studied during the last decades, due to the complexity of these systems defining their properties is extremely challenging. Using the experimental platform of cold atoms loaded into optical lattices, which has proven useful for controlled simulations of complex systems~\cite{review1, review2}, could open a new chapter for systematic studies of frustration. The simplest example of a frustrated system is that of spins on a triangular lattice with nearest neighbor anti-ferromagnetic couplings. Such geometric frustration was recently experimentally explored in a cold atom setting~\cite{seng1, seng2} by making use of spin degrees of freedom and engineered via external periodic driving. Experimental studies of frustrated systems, using cold atoms in optical lattices, are met with several difficulties. For example, without external driving, atoms in the lowest Bloch band typically support tunneling couplings with ferromagnetic order. Focused engineering of band structure degeneracies must go well beyond the tight-binding approximation, such that these are not lifted by higher order tunneling terms. Frustration may also appear on a square lattice due to different mechanisms. In the Villain model~\cite{villain}, three of the tunneling terms are positive, while the fourth one, closing the loop around a plaquette, is negative. Another possibility arises from the competition between nearest and next nearest neighbor tunneling terms. 

For classical rotors $s_{\bf i}=(\cos\phi_{\bf i},\sin\phi_{\bf i})$, the frustrated $XY$ model is~\cite{xy4a, xy4b}
\begin{equation}
\label{xy} 
H_{XY}=J_1\sum_{\langle{\bf ij}\rangle}\cos(\phi_{\bf i}-\phi_{\bf j})+J_2\sum_{\langle\!\langle{\bf ij}\rangle\!\rangle}\cos(\phi_{\bf i}-\phi_{\bf j}),
\end{equation}
with positive $J_1,\,J_2$ and the sums over nearest and next nearest neighbor coupling terms, respectively. The above model supports `interaction induced frustration' rather than the aforementioned geometric frustration. Despite four decades of intense research, the physics of the Villain and frustrated $XY$ models is not fully clarified. Both models, sharing the same universal features, support two symmetries; a continuous one, corresponding to particle conservation, and a chiral Ising-like symmetry. Due to the Mermin-Wagner theorem, the former cannot be spontaneously broken but nevertheless manifests as a Kosterlitz-Thouless transition. It is believed, but still debated~\cite{xy3a, xy3b, xy3c, xy3d, xy3e, xy3f, xy3g}, that upon lowering the temperature the models pass through two nearby transitions. The phase emerging between the transitions is argued to be a chiral spin liquid~\cite{xy8}. At zero temperature, and at the frustration point, $J_1=2J_2$, the ground state is a liquid phase. The liquid phase may also survive beyond the frustration point~\cite{xy7}, even though a thorough analysis, based on a large $N$-expansion, came to the conclusion that it exists only at the frustration point~\cite{xy6}. Spin wave theory suggests that for $J_2>2J_1$ anti-ferromagnetic order is established in the two sublattices with a relative `collinear' order~\cite{xy4a, xy4b}, as a result of the order-by-disorder phenomenon~\cite{od}. Needless to say, these speculations await any experimental verification.

\begin{figure}
\centerline{\includegraphics[width=8cm]{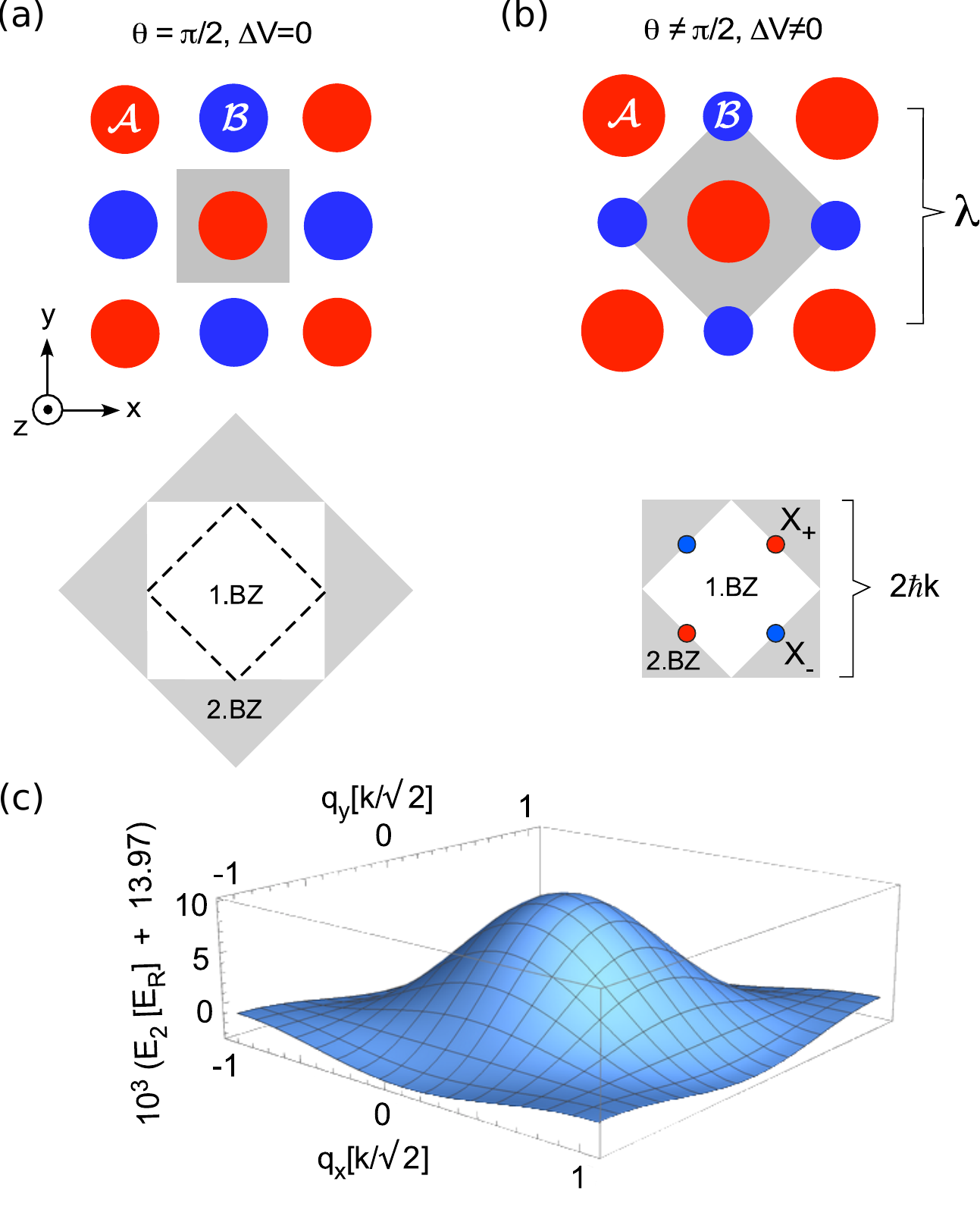}}
\caption{(Color online) (a) Conventional square lattice geometry with equivalent $\mathcal{A}$ and $\mathcal{B}$ sites, obtained for $\Delta V=0$ in Eq.~(\ref{pot}). The lower panel shows the first (white square) and second (grey triangles) BZs (dashed square denotes the reduced BZ for the distorted lattice). (b) Bipartite lattice geometry obtained for $\Delta V \neq 0$ in Eq.~(\ref{pot}) with the corresponding first (white square) and second (grey triangles) BZs in the lower panel. The two inequivalent high symmetry points $X_{\pm}$ are global minima of the second band. (c) The second band for $\Delta V\approx 0.063\,V_0$ and $V_0=10 E_\mathrm{R}$ is plotted across the first BZ.} 
\label{fig1}
\end{figure}

In this paper we propose, and experimentally realize, a surprisingly simple realization of the frustrated $XY$ model~(\ref{xy}) in terms of cold atoms loaded into an optical lattice. The advantage of our method is that no external driving (like lattice shaking or laser-induced tunneling) is required in order to realize the desired anti-ferromagnetic tunneling couplings and strong couplings beyond nearest neighbors. This becomes possible by considering meta-stable states in higher Bloch bands. In particular, we consider bosonic atoms loaded into the second band of a bipartite square lattice as described in Ref.~\cite{hemmerich}. This lattice provides two classes of alternating potential wells, denoted $\mathcal{A}$ and $\mathcal{B}$, with tunable relative depths $\Delta V$. In the case $\Delta V = 0$, a simple square lattice arises with a white square denoting the first Brillouin zone (BZ) in Fig.~\ref{fig1}(a), surrounded by four grey triangles showing the second BZ. The first Bloch band shows constant energy on the dashed rectangle in Fig.~\ref{fig1}(a), dividing the BZ into two equally sized areas. Tuning $\Delta V$ slightly away from zero has significant consequences. The unit cell area increases by a factor of two, i.e. the size of the first BZ shrinks by a factor two, such that it now is bordered by the dashed rectangle (Fig.~\ref{fig1}(b)). Hence, the lowest band of the $\Delta V = 0$ lattice splits into two new bands, thus giving rise to the first and second bands of the $\Delta V \neq 0$ lattice. If $\Delta V$ is kept close to zero, each of these bands is nearly degenerate along the first BZ edge. Thus, the second band provides an extended approximate minimum all along the edge of the first BZ just like for the frustrated $XY$ model (Fig.~\ref{fig1}(c)). At this stage our experiment is not capable of directly confirming the frustration emerging from such a quasi degeneracy. Nevertheless, we show in the following that numerical simulations of the atomic momentum distributions in the frustrated $XY$ model agree well with experimental time-of-flight measurements.

{\it Model system.} -- We consider bosonic atoms within a two-dimensional bipartite optical lattice. Along the transverse third direction the atoms are confined by a harmonic trap. The two-dimensional lattice is formed by superimposing two one-dimensional standing waves oscillating with a tunable relative phase $\theta$. The optical potential reads
\begin{equation}
\label{pot}
\begin{array}{lll}
V(x,y) &=& -V_0 \,[ \cos^2(kx)+\cos^2(ky)\\
& & +2\cos(\theta)\cos(kx)\cos(ky) ].
\end{array}
\end{equation}
In the remainder of the paper, energies are given in terms of the recoil energy $E_\mathrm{R}\equiv\hbar^2k^2/2m$, with $k$ the wave number and $m$ the atomic mass. 
For $\theta = \pi/2$, a conventional separable square lattice arises (cf. Fig.~\ref{fig1}(a)). When $\theta \neq \pi/2$ (cf. Fig.~\ref{fig1}(b)), the last term in Eq.~(\ref{pot}) distorts the lattice. This gives rise to a bipartite non-separable lattice geometry, composed of two sub-lattices, with deep and shallow sites $\mathcal{A}$ and $\mathcal{B}$, respectively. Their potential energy difference is $\Delta V=-4V_{0}\cos(\theta)$. 
Due to this distortion, the first BZ is decreased by a factor of two, and the first Bloch band of the $\theta = \pi/2$ lattice splits into two new bands defining a new BZ. These bands are separated by a gap, which for weak distortions scales as $\sim \Delta V / V_0$ according to the `nearly free electron gas' model~\cite{ashcroft1976solid}. It is striking how flat the bands are along the BZ edge as compared to the band gap. Reaching the regime of frustration requires a flatness characterized by an energy spread which is below the energy associated with two-body contact interaction between the atoms. Increasing the flatness, however, comes with the price that the band width and the band gap are decreased. Decreasing the band width slows down the relaxation toward the regime of frustration, while decreasing the band gap increases the inter-band relaxation, i.e., the loss of atoms from the second band via collisions~\cite{metastable}. Thus, there is a trade-off in choosing the optimal configuration. For the rest of the paper we pick parameters $V_0=10\,E_\mathrm{R}$ and $\Delta V=0.063\,V_0$, for which we plot the second band in fig.~\ref{fig1}(c). The flatness is not perfect; the band width is about six times the energy spread along the BZ edge, while the gap is roughly 230 times this spread. 

To explore the properties of atoms residing solely in the second band, we evoke a single-band approximation and derive an effective lattice model,
\begin{equation}
\hat H_{1}=\sum_{{\bf i}\neq{\bf j}}(t_{\bf ij}\hat a_{\bf i}^\dagger\hat a_{\bf j}+h.c.),
\end{equation}
where the matrix $t_{\bf ij}$ amounts to the tunneling amplitudes between sites ${\bf i}$ and ${\bf j}$. These amplitudes can be chosen real. The above lattice Hamiltonian becomes diagonal by Fourier-transforming it. Using this we can identify $t_{\bf ij}$ as the Fourier coefficients of the dispersion $E_2(q_x,q_y)$ for the second band of the full Hamiltonian. The relevant numerically extracted amplitudes are presented in tab.~\ref{tab1}. For the present parameters, it is found that the system is close to the frustrated scenario, i.e. after identifying $J_1$ and $J_2$ we have $J_1/J_2=1.43$ while other tunneling terms are approximately zero in comparison. The ratio $J_1/J_2$ does not equal the desired value 2, implying that the second band is not fully flat along the BZ edge, as also seen in fig.~\ref{fig1}(c). The final step in mapping to the frustrated $XY$ model is accomplished by considering the mean-field approximation in which the atomic operators are replaced by their expectations, e.g. $\hat a\rightarrow\alpha_{\bf i}=\sqrt{n_{\bf i}}e^{i\phi_{\bf i}}$ such that $n_{\bf i}$ is the (average) number of atoms at site ${\bf i}$ and $\phi_{\bf i}$ is the intra-site condensate phase.

\begin{table}[ht]
\begin{tabular}{|c|c|c|c|}
\hline
\hspace{0.2cm}$t_{nm}$\hspace{0.2cm} & \hspace{0.65cm}0 \hspace{0.65cm} &  \hspace{0.65cm}1\hspace{0.65cm} & \hspace{0.65cm}2 \hspace{0.65cm} \\
\hline\hline
0 & -13.97 & $1.4\cdot10^{-3}$& $-6\cdot10^{-6}$\\
\hline
1 & $1.4\cdot10^{-3}$& $9.7\cdot10^{-4}$& $-4\cdot10^{-6}$\\
\hline
2 & $-6\cdot10^{-6}$ &$-4\cdot10^{-6}$ & $-6\cdot10^{-7}$\\
\hline
\end{tabular}
\caption{Tunneling matrix for the parameters $V_0=10 E_\mathrm{R}$ and $\Delta V=0.063V_{0}$. The first diagonal term gives the onsite energy, the terms $(0,1)$ and $(1,0)$ are the nearest neighbor tunneling rates $J_1$, and $(1,1)$ the next nearest neighbor tunneling rate $J_2$ and so on.}
\label{tab1}
\end{table}

\begin{figure}
\centerline{\includegraphics[width=8cm]{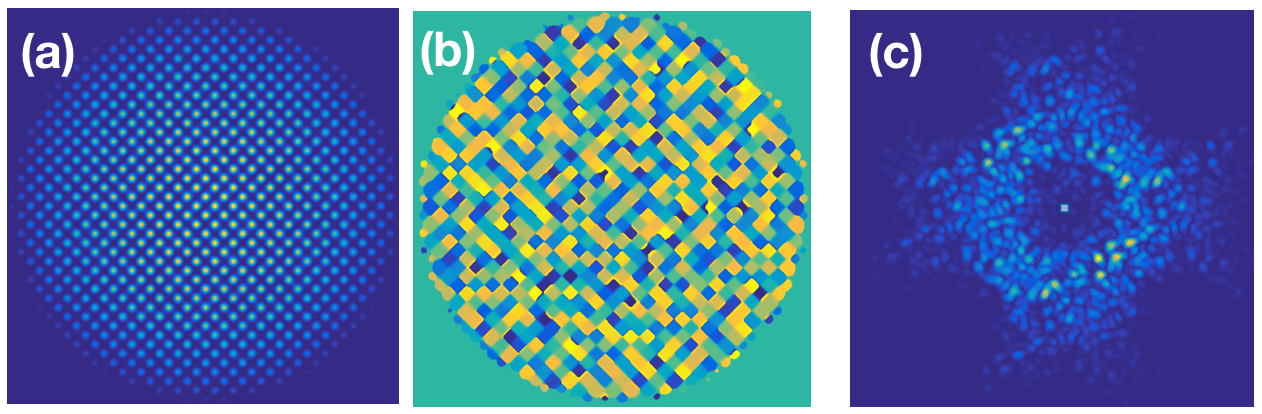}}
\caption{(Color online) Numerical simulation of the atomic real space density (a), its phase (b), and the momentum density profile (c). The relaxation is mimicked by imaginary time-propagation for a dimensionless time $t_f=100$. At this stage the condensate is still fairly uniformly distributed around the BZ edge (c) The middle plot (b) shows how no evident order is established in the condensate, which signals the presence of frustration. The system parameters are chosen to represent the experimental ones; the dimensionless trapping frequency $\omega=0.037$, $V_0=10E_R$, $\Delta V=0.063\,V_{0}$, $U_0=0.011E_R$ and the number of atoms $N=10^5$. } 
\label{fig2}
\end{figure}

We must further add one term accounting for atom-atom interaction and another corresponding to the confining harmonic trap. Following the standard assumptions, i.e. the local density approximation and onsite $s$-wave scattering, we then derive the mean-field Hamiltonian
\begin{equation}
H=H_1+\sum_{\bf i}\frac{\omega^2}{2}n_{\bf i}|{\bf i}|^2+\frac{U}{2}\sum_{\bf i}n_{\bf i}^2,
\end{equation}
where $\omega$ and $U$ are the dimensionless trap frequency and interaction strength respectively. We can simulate the atomic relaxation that occurs in the experiment (see below) by initializing a state $\psi({\bf i},0)$, defined by some set of random variables $n_{\bf i}$ and $\phi_{\bf i}$, and then evolving this state in imaginary time~\cite{imagt}. If the system would not be frustrated, the initial lack of long range order rapidly disappears and a phase-locking among the $\phi_{\bf i}$ will occur. For example, if the BZ edge is not flat enough (with minima at the $X_\pm$ points), the global condensate builds up a striped order characterized by a $\pi$-phase difference between either the columns or rows of the lattice. However, if the BZ is approximately flat, such order will not build up on any experimentally relevant time scale. In Fig.~\ref{fig2} we show properties of the numerically extracted condensate. Here, at a given time $t$, the condensate order parameter is obtained from $\Psi(x,y)=\sum_{\bf i}\alpha_{\bf i}w_{\bf i}(x,y)$, where $w_{\bf i}(x,y)$ is the Wannier function at site ${\bf i}$, which for the figure has been taken as a real Gaussian. From $\Psi(x,y)$ we get the atomic and momentum densities $\rho(x,y)$ ($=|\Psi(x,y)|^2$) and $\tilde\rho(p_x,p_y)$ respectively. We note how the relaxation causes the momentum distribution (c) to assemble along the BZ edge. The spatial density (a) shows an envelope following the inverse shape of the trap with seemingly complex local density fluctuations stemming from the frustration. A manifestation of the frustration is especially evident in (b) where we plot the condensate phase $\phi(x,y)=\mathrm{angle}[\Psi(x,y)]$. If we were to let the condensate relax further, i.e. by propagating the simulation further, the momentum distribution would get more and more concentrated around the $X_\pm$ points, and domains of striped order would form in the spatial condensate density. The sizes of these domains would grow with time to eventually suppress the frustration completely. 

{\it Experimental demonstration.} -- A nearly pure spin-polarised Bose-Einstein Condensate of $3 \cdot 10^{5}$ rubidium ($^{87}$Rb) atoms in the hyperfine state $|F=2,m_F=2\rangle$ with a temperature of $30\,$nK is produced in a crossed optical dipole trap (ODT). The atomic sample is adiabatically loaded into a bipartite optical square lattice, which displays a double-well structure as described in eq.~(\ref{pot}). The optical lattice is formed by two mutually orthogonal optical standing waves with a wavelength of $\lambda =1064\,$nm. The lattice depth $V_{0}$ as well as the potential energy difference $\Delta V$ can be actively tuned. The tuning of $\Delta V$ is realized via the control of the time phase difference $\theta$ between the standing waves with a precision of $\pi/300$ conforming to $\Delta V=-4V_{0}\cos(\theta)$. Rapid adjustment of $\Delta V$ allows us to quench the atoms into higher Bloch bands. For technical details see refs.~\cite{fermionlattice, bosonlattice}.

\begin{figure}
\centerline{\includegraphics[width=9.2cm]{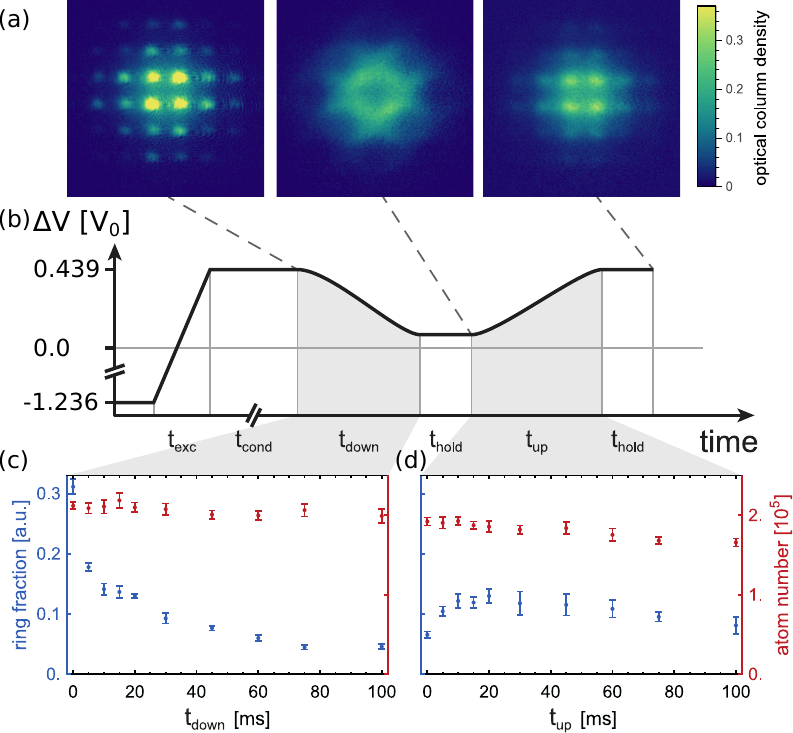}}
\caption{Collapse and revival of coherence. Three momentum spectra are shown in (a) for characteristic values of $\Delta V$ tuned according to the protocol in (b). The first image at $\Delta V=0.439\,V_{0}$ shows coherent condensate fractions at the $X_\pm$ points of the second band (cf. Fig. \ref{fig1}(b)). In the second image at $\Delta V=0.063\,V_{0}$ coherence has completely vanished. The third image, recorded after ramping back to $\Delta V=0.439\,V_{0}$, shows partially restored coherence. The blue symbols in (c) and (d) show the observed collapse and revival of coherent condensate fractions at the $X_\pm$ points plotted against the ramp times $t_{\rm down}$ and $t_{\rm up}$, respectively. The red symbols indicate the total population of the first and second band. Error bars correspond to the standard deviation of the mean for a set of eight measurements.}  
\label{fig3}
\end{figure}

The experimental sequence is performed as follows (cf. Fig.~\ref{fig3}): the atoms are initially loaded into the lowest band of the lattice by adiabatically ramping up the lattice depth $V_{0}$ to $10\,E_{\rm {R}}$ in $150\,{\rm ms}$ with fixed $\Delta V = -1.236\,V_{0}$. In addition, we are able to reduce the temperature of the atomic sample by simultaneously decreasing the potential depth of the ODT. After a hold time of $100\,$ms, a quench protocol is applied to excite the atoms into a metastable second Bloch band~\cite{metastable} by rapidly tuning $\Delta V$ to $0.439\,V_{0}$ in $t_{\rm exc} = 0.1\,$ms. After a holding time of $t_{\rm cond} = 500\,$ms, the atoms are condensed at the two inequivalent energy minima of the band, i.e. the $X_{\pm}$-points (cf. Fig.~\ref{fig3} (a) left). Subsequently, the potential difference $\Delta V$ is ramped down during a variable time $t_{{\rm down}}\in\{0,100\}\,$ms to $\Delta V=0.063\,V_{0}$. After $t_{{\rm hold}}= 0.02\,{\rm ms}$ the potential is ramped back to $\Delta V=0.439\,V_{0}$ in a variable time $t_{up}\in\{0,100\}\,{\rm ms}$ for a fixed $t_{{\rm down}}$ and the atoms are held in the lattice for another $t_{{\rm hold}}= 0.02\,{\rm ms}$. Finally, the lattice and trapping potentials are instantaneously switched off followed by a ballistic expansion of $22\,{\rm ms}$ and the momentum spectrum is recorded via an absorption image. The temporal evolution of the coherent fraction, $\left(n_{\rm coh}-n_{\rm incoh}\right)/\left(n_{\rm coh}+n_{\rm incoh}\right)$ of the population at the $X_\pm$ points is observed. This observable is determined by counting the atoms in a disk-shaped region of interest (ROI) around both $X_{\pm}$ points $n_{\rm coh}$ and subtracting the atoms $n_{\rm incoh}$ enclosed by a ring-shaped ROI with the same area of the disk-shaped ROIs~\cite{metastable}.

Figure~\ref{fig3}(a) shows momentum spectra at $\Delta V=0.439\,V_{0}$ before the down-ramp (left), at $\Delta V=0.063\,V_{0}$ after the $t_{\rm down}=60\,$ms down-ramp and a short hold time $t_{\rm hold}=20\,\mu$s (centre), and after the $t_{up}=20\,$ms up-ramp back to $\Delta V=0.439\,V_{0}$ followed by $t_{\rm hold}=20\,\mu$s (right). At $\Delta V = 0.439\,V_{0}$, the second band possesses distinct energy minima at the $X_\pm$ points and a coherent condensate fraction is observed as indicated by the clearly visible higher order Bragg resonances. When the potential difference is ramped down to $\Delta V=0.063 V_{0}$ (cf. Fig.~\ref{fig3} (b)), the condensate fraction decreases. As illustrated by the blue data points in Fig.~\ref{fig3}(c), if the down-ramp time $t_{{\rm down}}$ is sufficiently large ($> 60\,$ms) to satisfy adiabaticity, the condensate fraction practically vanishes and the atomic sample no longer displays coherence. At $\Delta V=0.063V_{0}$, the energy-momentum dispersion of the second band is almost degenerate along the edge of the first Brillouin zone. The atoms are observed to spread over the entire edge of the second Bloch band populating all available Bloch states. The associated momentum distribution in the centre image of Fig.~\ref{fig3}(a) shows a rectangular shape, similar as in the theoretical prediction of Fig.~\ref{fig2}(c). The loss of coherence indicates that the condensate suffers from excessive fragmentation due to the large number of nearly degenerate minimal energy states.

In order to exclude the possibility that the observed vanishing of coherence is due to uncontrolled heating, we have added a second stage in the experimental protocol in Fig.~\ref{fig3}(b), where $\Delta V$ is ramped up again to the initial value $\Delta V=0.439\,V_{0}$. If this ramp is processed adiabatically, we observe a revival of the condensate fraction, as is seen in the right image in Fig.~\ref{fig3}(a). Figure~\ref{fig3}(d), shows that in fact the recovery of the condensed fraction at the $X_\pm$ points requires a sufficiently long ramp time $t_{{\rm up}}$ to satisfy adiabaticity. Maximum coherence is found for $t_{{\rm up}}$ around $20\,$ms. The total number of atoms in the first and second band remains nearly constant throughout the measurement (red markers in Fig.~\ref{fig3}(c) and (d)). The observed recovery of the condensate fraction appears incomplete, possibly due to residual band relaxation at $\Delta V=0.063 V_{0}$ and associated heating processes. These findings indicate that the observed loss of coherence, when tuning to $\Delta V=0.063V_{0}$, could be interpreted as a signature of a glassy phase originating from massive degeneracy of the second band. A similar protocol was used in Ref.~\cite{Gre:02} to discriminate a Mott insulation phase from a thermally incoherent phase.

{\it Conclusion.} -- We proposed a new method, relying on a lattice distortion, for realizing frustrated models in an optical lattice system. Our proposal does not rely on any driving as in earlier schemes, but the desired anti-ferromagnetic tunneling terms result naturally by considering the first excited band. The method was experimentally demonstrated by loading a gas of ultracold atoms to this second band, and the condensate coherence fraction was measured upon tuning from a non-frustrated to a frustrated regime and then back. In the frustrated regime no visible signatures of coherence were observed, in agreement with numerical simulations of the frustrated system, while coherence was reestablished after ramping back to the non-frustrated regime. In a strict sense, this is no direct evidence of frustration, which would require different experimental tools, e.g. single-site pair-correlation measurements could give more insight into possible glass and liquid phases. Nevertheless, our work is a first step realizing systematic experimental studies of frustrated matter. 

\begin{acknowledgements}
We thank Axel Gagge and Themistoklis Mavrogordatos for helpful discussions. We acknowledge financial support from the Knut and Alice Wallenberg foundation and the Swedish research council (VR). We acknowledge support from the Deutsche Forschungsgemeinschaft (DFG) through the collaborative research center SFB 925 (Project No. 170620586, C1).  
\end{acknowledgements}

\end{document}